\RequirePackage{lineno}
\documentclass[aps,twocolumn,superscriptaddress,preprintnumbers,amsmath,amssymb,floatfix,nofootinbib]{revtex4}

\usepackage{graphicx} 
\usepackage{dcolumn}  
\usepackage{rotating} 
\usepackage{lscape}

\usepackage{bm}
\usepackage{url} 
\usepackage{amsmath} 
\usepackage{amssymb} 

\usepackage{verbatim}
\usepackage{multirow}
\usepackage{subfigure}

\graphicspath{{ps}}

\begin{document}
%\linenumbers

\title{Axion Dark Matter Search around 4.55 $\mu$eV with Dine-Fischler-Srednicki-Zhitnitskii Sensitivity}

\author{Andrew K. Yi}\affiliation{Dept.\ of Physics, Korea Advanced Institute of Science and Technology, Daejeon 34141, Republic of Korea}\affiliation{Center for Axion and Precision Physics Research, Institute for Basic Science, Daejeon 34051, Republic of Korea}
\author{Saebyeok Ahn}\affiliation{Dept.\ of Physics, Korea Advanced Institute of Science and Technology, Daejeon 34141, Republic of Korea}\affiliation{Center for Axion and Precision Physics Research, Institute for Basic Science, Daejeon 34051, Republic of Korea}
\author{\c{C}a\u{g}lar Kutlu}\affiliation{Dept.\ of Physics, Korea Advanced Institute of Science and Technology, Daejeon 34141, Republic of Korea}\affiliation{Center for Axion and Precision Physics Research, Institute for Basic Science, Daejeon 34051, Republic of Korea}
\author{JinMyeong Kim}\affiliation{Dept.\ of Physics, Korea Advanced Institute of Science and Technology, Daejeon 34141, Republic of Korea}\affiliation{Center for Axion and Precision Physics Research, Institute for Basic Science, Daejeon 34051, Republic of Korea}
\author{Byeong Rok Ko}\email[Corresponding author~:~]{brko@ibs.re.kr}\affiliation{Center for Axion and Precision Physics Research, Institute for Basic Science, Daejeon 34051, Republic of Korea}
\author{Boris I. Ivanov}\affiliation{Center for Axion and Precision Physics Research, Institute for Basic Science, Daejeon 34051, Republic of Korea}
\author{HeeSu Byun}\affiliation{Center for Axion and Precision Physics Research, Institute for Basic Science, Daejeon 34051, Republic of Korea}
\author{Arjan F. van Loo}\affiliation{RIKEN Center for Quantum Computing (RQC), Wako, Saitama 351-0198, Japan}\affiliation{Dept.\ of Applied Physics, Graduate School of Engineering, The University of Tokyo, Bunkyo-ku, Tokyo 113-8656, Japan}
\author{SeongTae Park}\affiliation{Center for Axion and Precision Physics Research, Institute for Basic Science, Daejeon 34051, Republic of Korea}
\author{Junu Jeong}\affiliation{Center for Axion and Precision Physics Research, Institute for Basic Science, Daejeon 34051, Republic of Korea}
\author{Ohjoon Kwon}\affiliation{Center for Axion and Precision Physics Research, Institute for Basic Science, Daejeon 34051, Republic of Korea}

\author{Yasunobu Nakamura}\affiliation{RIKEN Center for Quantum Computing (RQC), Wako, Saitama 351-0198, Japan}\affiliation{Dept.\ of Applied Physics, Graduate School of Engineering, The University of Tokyo, Bunkyo-ku, Tokyo 113-8656, Japan}
\author{Sergey V. Uchaikin}\affiliation{Center for Axion and Precision Physics Research, Institute for Basic Science, Daejeon 34051, Republic of Korea}
\author{Jihoon Choi}\altaffiliation[]{Now at Korea Astronomy and Space Science Institute, Daejeon 34055, Republic of Korea}\affiliation{Center for Axion and Precision Physics Research, Institute for Basic Science, Daejeon 34051, Republic of Korea}
\author{Soohyung Lee}\affiliation{Center for Axion and Precision Physics Research, Institute for Basic Science, Daejeon 34051, Republic of Korea}
\author{MyeongJae Lee}\altaffiliation[]{Now at Dept.\ of Physics, Sungkyunkwan University, Suwon 16419, Republic of Korea}\affiliation{Center for Axion and Precision Physics Research, Institute for Basic Science, Daejeon 34051, Republic of Korea}
\author{Yun Chang Shin}\affiliation{Center for Axion and Precision Physics Research, Institute for Basic Science, Daejeon 34051, Republic of Korea}
\author{Jinsu Kim}\affiliation{Dept.\ of Physics, Korea Advanced Institute of Science and Technology, Daejeon 34141, Republic of Korea}\affiliation{Center for Axion and Precision Physics Research, Institute for Basic Science, Daejeon 34051, Republic of Korea}
\author{Doyu Lee}\altaffiliation[]{Now at Samsung Electronics, Gyeonggi-do 16677, Republic of Korea}\affiliation{Center for Axion and Precision Physics Research, Institute for Basic Science, Daejeon 34051, Republic of Korea}
\author{Danho Ahn}\affiliation{Dept.\ of Physics, Korea Advanced Institute of Science and Technology, Daejeon 34141, Republic of Korea}\affiliation{Center for Axion and Precision Physics Research, Institute for Basic Science, Daejeon 34051, Republic of Korea}
\author{SungJae Bae}\affiliation{Dept.\ of Physics, Korea Advanced Institute of Science and Technology, Daejeon 34141, Republic of Korea}\affiliation{Center for Axion and Precision Physics Research, Institute for Basic Science, Daejeon 34051, Republic of Korea}
\author{Jiwon Lee}\affiliation{Dept.\ of Physics, Korea Advanced Institute of Science and Technology, Daejeon 34141, Republic of Korea}\affiliation{Center for Axion and Precision Physics Research, Institute for Basic Science, Daejeon 34051, Republic of Korea}
\author{Younggeun Kim}\affiliation{Center for Axion and Precision Physics Research, Institute for Basic Science, Daejeon 34051, Republic of Korea}
\author{Violeta Gkika}\affiliation{Center for Axion and Precision Physics Research, Institute for Basic Science, Daejeon 34051, Republic of Korea}
\author{Ki Woong Lee}\affiliation{Center for Axion and Precision Physics Research, Institute for Basic Science, Daejeon 34051, Republic of Korea}
\author{Seonjeong Oh}\affiliation{Center for Axion and Precision Physics Research, Institute for Basic Science, Daejeon 34051, Republic of Korea}
\author{Taehyeon Seong}\affiliation{Center for Axion and Precision Physics Research, Institute for Basic Science, Daejeon 34051, Republic of Korea}
\author{DongMin Kim}\affiliation{Center for Axion and Precision Physics Research, Institute for Basic Science, Daejeon 34051, Republic of Korea}
\author{Woohyun Chung}\affiliation{Center for Axion and Precision Physics Research, Institute for Basic Science, Daejeon 34051, Republic of Korea}
\author{Andrei Matlashov}\affiliation{Center for Axion and Precision Physics Research, Institute for Basic Science, Daejeon 34051, Republic of Korea}
\author{SungWoo Youn}\affiliation{Center for Axion and Precision Physics Research, Institute for Basic Science, Daejeon 34051, Republic of Korea}
\author{Yannis K. Semertzidis}\affiliation{Center for Axion and Precision Physics Research, Institute for Basic Science, Daejeon 34051, Republic of Korea}\affiliation{Dept.\ of Physics, Korea Advanced Institute of Science and Technology, Daejeon 34141, Republic of Korea}

\begin{abstract}
  We report an axion dark matter search at
  Dine-Fischler-Srednicki-Zhitnitskii sensitivity with the CAPP-12TB
  haloscope, assuming axions contribute 100\% of the local dark matter
  density.
  The search excluded the axion--photon coupling
  $g_{a\gamma\gamma}$ down to about $6.2\times10^{-16}$ GeV$^{-1}$ over
  the axion mass range between 4.51 and 4.59 $\mu$eV at a 90\%
  confidence level.  
  The achieved experimental sensitivity can also exclude
  Kim-Shifman-Vainshtein-Zakharov axion dark matter that makes up just
  13\% of the local dark matter density.  
  The CAPP-12TB haloscope will continue the search over a wide range
  of axion masses.  
\end{abstract}
\maketitle
\tighten

{\renewcommand{\thefootnote}{\fnsymbol{footnote}}}
\setcounter{footnote}{0}

The standard model of Big Bang cosmology combined with precision
cosmological measurements strongly suggest that cold dark matter
(CDM) constitutes about 85\% of the matter and 27\% of the energy
density in the Universe~\cite{PLANCK}.
CDM is a subject of beyond the standard model of particle physics (SM)
and remains hidden to date.
The axion~\cite{AXION} stems from the breakdown of a new global
symmetry introduced by Peccei and Quinn~\cite{PQ} to solve the strong
$CP$ problem in the SM~\cite{strongCP}, and is one of the most
prominent CDM candidates, provided its mass is above
$\mathcal{O}(\mu$eV) according to the original work~\cite{CDM_LOW}, or
above $\mathcal{O}$(peV) by more recent works~\cite{AXION_PROD1}, and
below $\mathcal{O}$(meV)~\cite{AXION_PROD4,SN1987}.

The axion haloscope search proposed by Sikivie~\cite{sikivie} exploits
the axion--photon coupling $g_{a\gamma\gamma}$ in a microwave cavity
permeated with a static magnetic field, which results in resonant
conversions of axions to photons when the axion mass $m_a$ matches the
frequency of the cavity mode $\nu$, $m_a=h\nu/c^2$.
The two most popular models, Kim-Shifman-Vainshtein-Zakharov
(KSVZ)~\cite{KSVZ} and Dine-Fischler-Srednicki-Zhitnitskii
(DFSZ)~\cite{DFSZ}, benchmark the $g_{a\gamma\gamma}$ with
$g_\gamma=-0.97$ and 0.36, respectively, where $g_\gamma$ is a
dimensionless coupling constant and comes from
$g_{a\gamma\gamma}=\frac{\alpha g_{\gamma}}{\pi f_a}$ along with the
axion decay constant $f_a$ and the fine structure constant $\alpha$.
The use of a high quality microwave cavity makes the axion haloscope
the most promising method for axion dark matter searches in the
microwave region.

In this Letter, we report a DFSZ axion dark matter search using the
CAPP-12TB haloscope at the Institute for Basic Science (IBS) Center
for Axion and Precision Physics Research (CAPP)~\cite{CAPP,8TB_PRL}.
Here, 12TB stands for our solenoid specifications, the central
magnetic field of 12~T and the Big bore of 320~mm~\cite{OI}. To date,
DFSZ axion dark matter sensitive searches were only achieved by the
Axion Dark Matter eXperiment (ADMX)~\cite{ADMX_DFSZ1, ADMX_DFSZ2}.
\begin{figure}[h]
  \centering
  \includegraphics[width=0.42\textwidth]{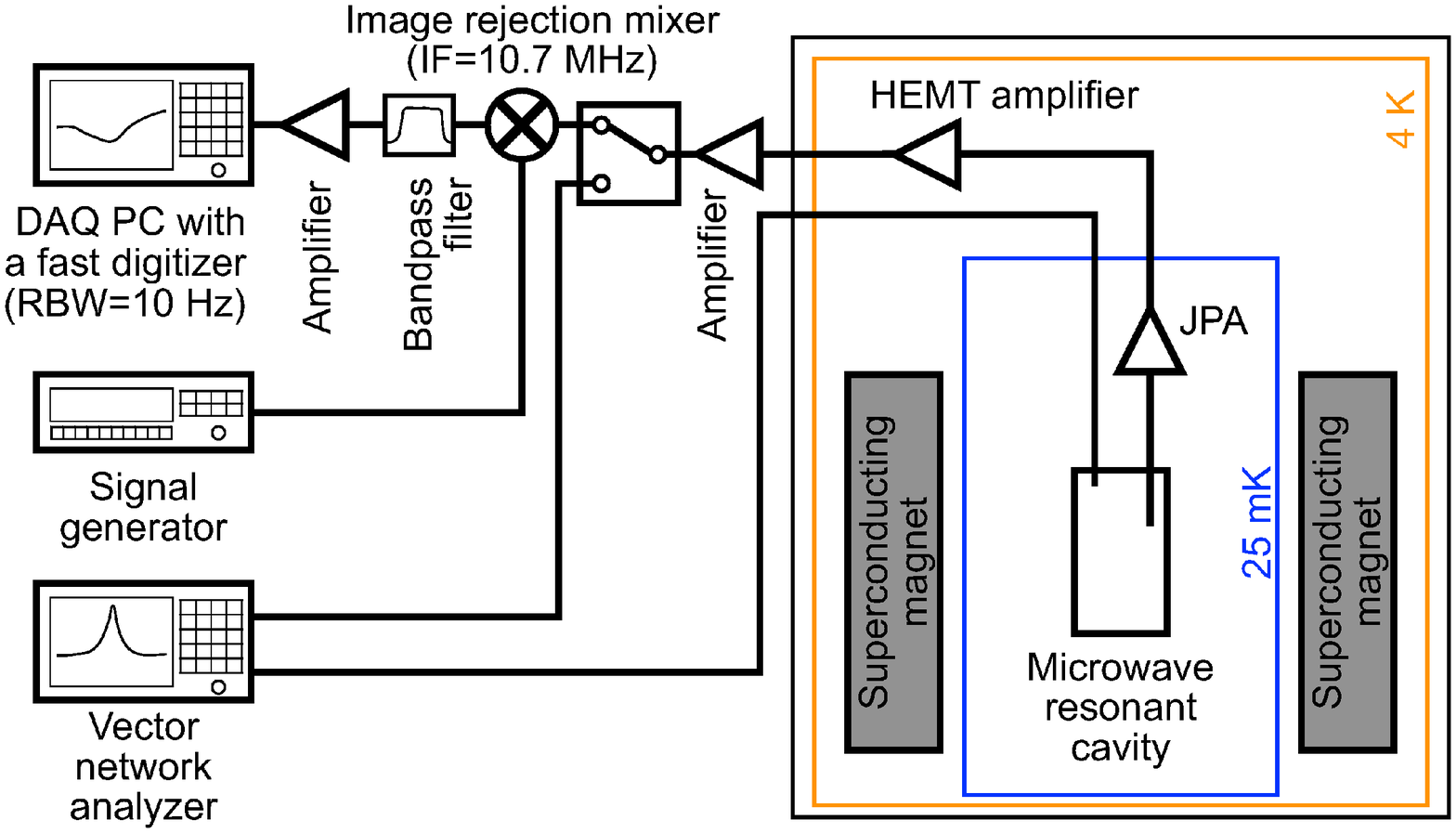}
  \caption{Schematic of the CAPP-12TB haloscope.}
  \label{FIG:CAPP-12TB}
\end{figure}

The CAPP-12TB haloscope depicted in Fig.~\ref{FIG:CAPP-12TB} comprises
a 36.85~L frequency tunable copper cylindrical cavity placed at the
magnet center, a superconducting solenoid whose rms magnetic field
$B_{\rm rms}$ over the cavity volume $V$ is 10.31~T, and a heterodyne
receiver chain with a Josephson Parametric Amplifier (JPA) as the
first amplifier. The experiment maintained the physical temperatures
of the cavity and the JPA at around 25~mK using a dilution
refrigerator DRS-1000~\cite{DRS-1000} whose cooling power was measured
to be about 1~mW at 90~mK without any load.

The detected axion signal power is expected to be
\begin{eqnarray}
  P^{a\gamma\gamma}_{a}&=&22.51~{\rm yW}
  \left(\frac{g_\gamma}{0.36}\right)^2
  \left(\frac{B_{\rm rms}}{10.31~{\rm T}}\right)^2\nonumber \\
  &\times&\left(\frac{V}{36.85~{\rm L}}\right) 
  \left(\frac{C}{0.6}\right) 
  \left(\frac{Q_L}{35000}\right)\nonumber \\
  &\times&\left(\frac{\nu}{1.1~{\rm GHz}}\right)
  \left(\frac{\rho_a}{0.45~{\rm GeV/cm^{3}}}\right) 
  \label{EQ:PAXION}
\end{eqnarray}
when the axion mass matches the frequency of the cavity mode
$m_a=h\nu/c^2$ and the cavity mode coupling to the receiver, $\beta$,
is 2. In Eq.~(\ref{EQ:PAXION}), $C$ is the cavity-mode-dependent form
factor which in practice includes the axion--photon interaction energy
normalized by the energy stored in the electric and magnetic fields over
the cavity volume~\cite{EMFF_BRKO}, $Q_L$ is the loaded quality factor
of the cavity mode, and $\rho_a$ is the local dark matter density.
Here, we have assumed axions make up 100\% of the local dark matter
density, i.e., $\rho_a=0.45$ GeV/cm$^3$, to explore the
$g_{a\gamma\gamma}$. As the CAPP-12TB axion haloscope has cylindrical
geometry, the chosen cavity mode for this search is the
TM$_{010}$-like mode, to maximize $C$. Assuming the standard halo
model of axion dark matter, the signal power given in
Eq.~(\ref{EQ:PAXION}) would then be distributed over a boosted
Maxwellian shape with an axion rms speed of about 270~km/s, and the
Earth rms speed of 230~km/s with respect to the galaxy
frame~\cite{AXION_SHAPE}, respectively, which is the model for this
work.

\begin{figure}
  \centering
  \includegraphics[width=0.37\textwidth]{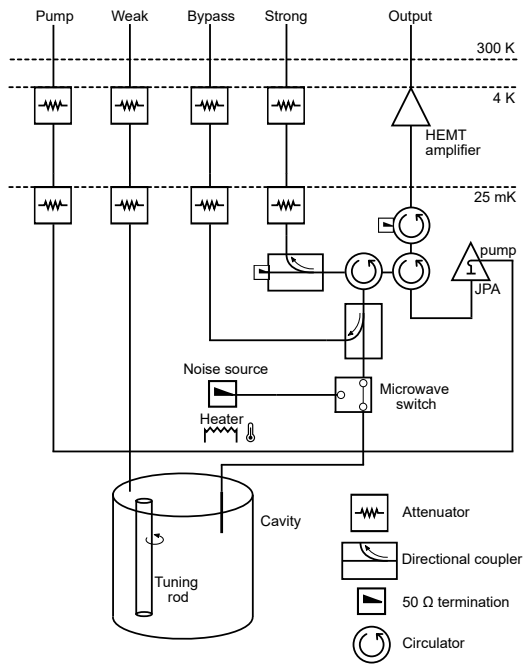}  
  \caption{CAPP-12TB receiver diagram.}
  \label{FIG:CAPP-12TB-chain}
\end{figure}

We have realized a frequency tuning mechanism operating in a
cryogenic and high magnetic field environment employing a
piezoelectric motor manufactured by attocube~\cite{ATTOCUBE}. The
motor sits on the top end cap of the cavity and links the tuning rod
directly through a crank arm, where the tuning rod is a copper
cylinder whose diameter is about a tenth of the cavity diameter.
By rotating the tuning rod about the tuning axle, we tuned the cavity
modes over the frequency range considered in this work.
Another attocube piezoelectric motor for an antenna has been adopted
to adjust $\beta$ that was measured to be 1.8--2.2 during data taking
through the ``Strong'' line in Fig.~\ref{FIG:CAPP-12TB-chain}.
A fixed antenna minimally coupled to the cavity has also been
implemented at the end of the ``Weak'' line.
The $Q_L$ of the TM$_{010}$-like mode for each frequency step
was measured to be 35000--38000 and the relevant unloaded cavity
quality factor $Q_0=Q_L(1+\beta)$ was 107000--114000, and the
mode-dependent $C$ were calculated to be about 0.6 using a finite
element method calculation~\cite{CSTCOMSOL}.
\begin{figure*}
  \centering
  \subfigure{\includegraphics[width=0.3\textwidth]{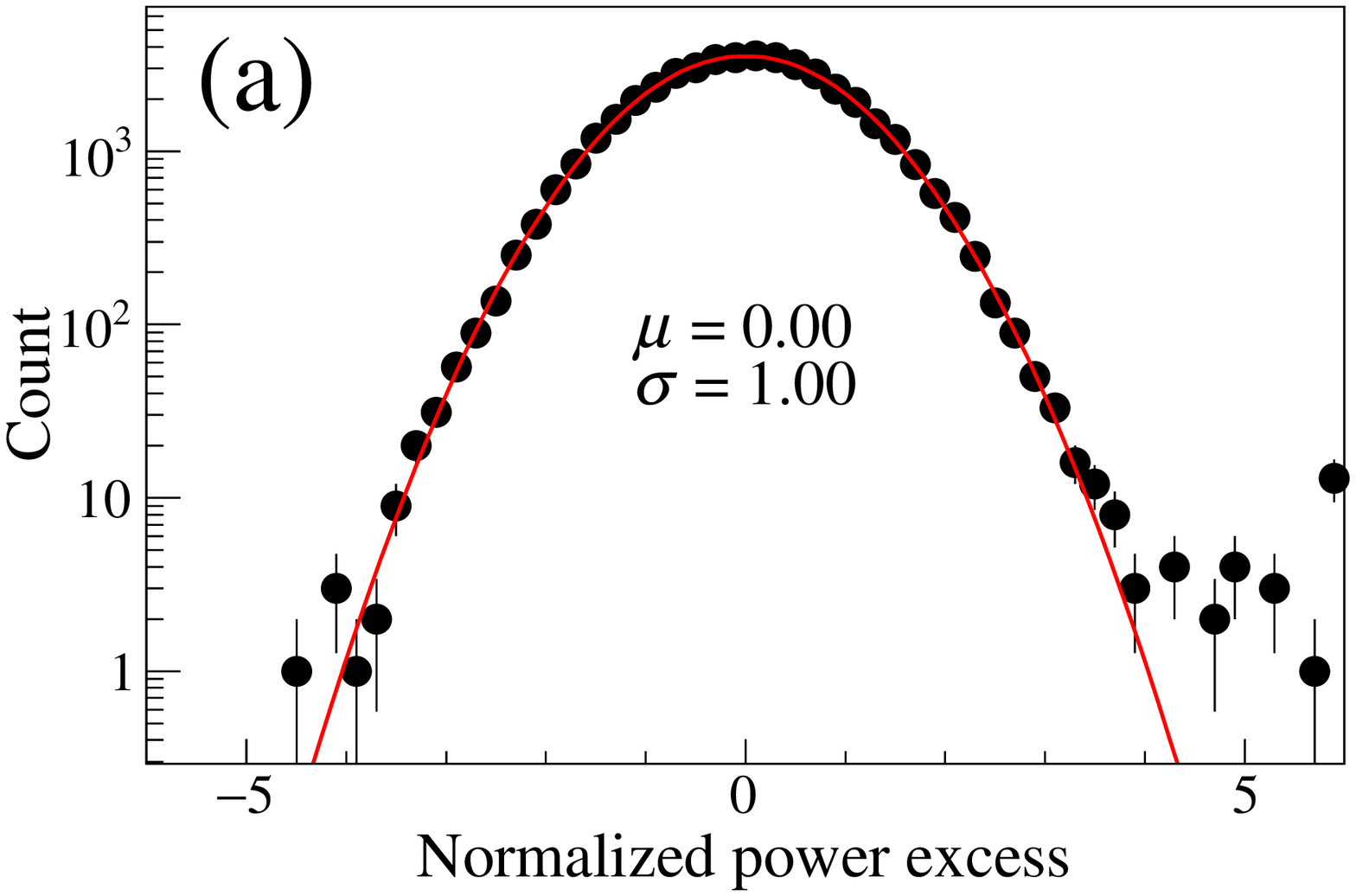}}
  \subfigure{\includegraphics[width=0.3\textwidth]{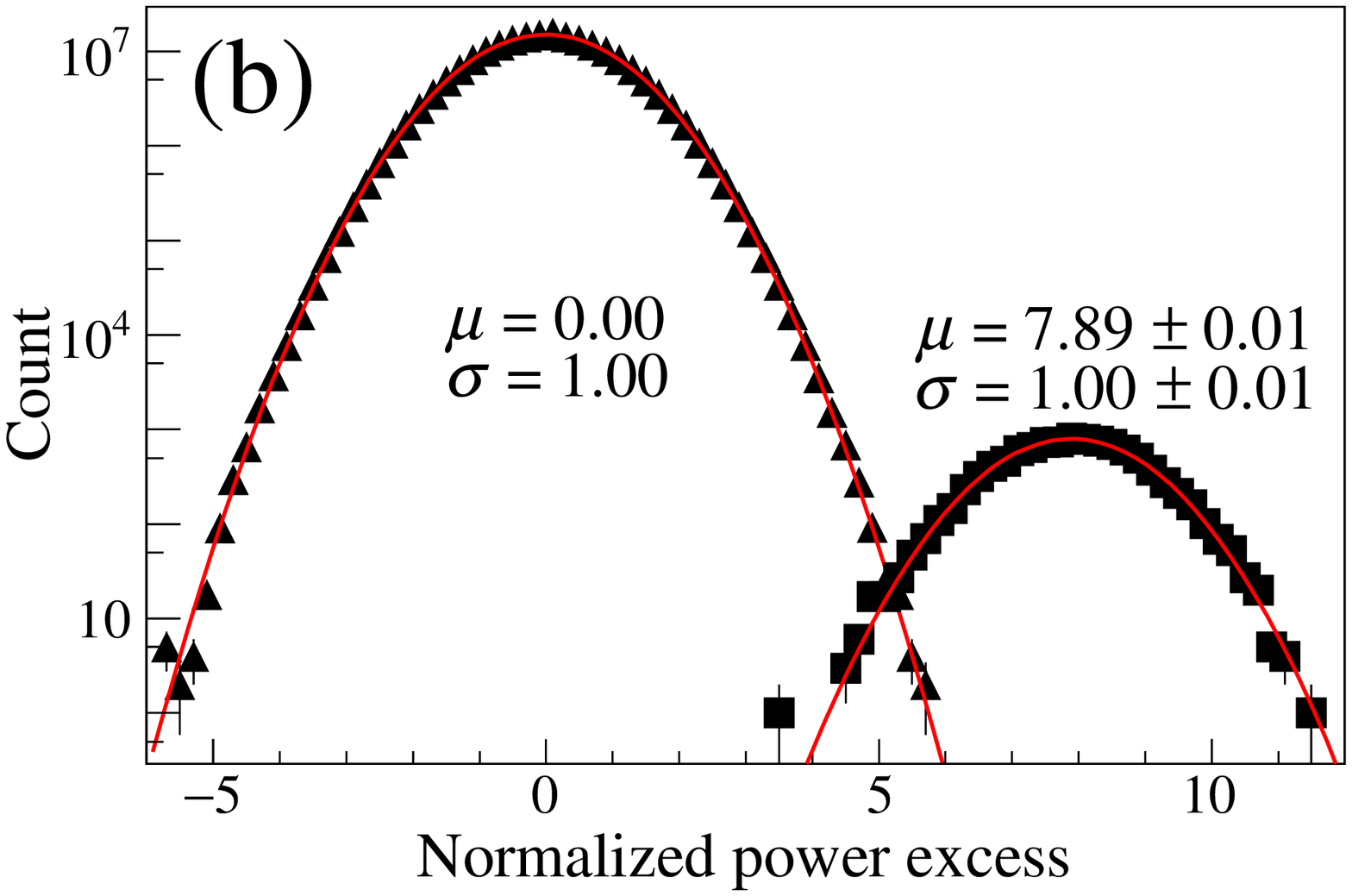}}
  \subfigure{\includegraphics[width=0.3\textwidth]{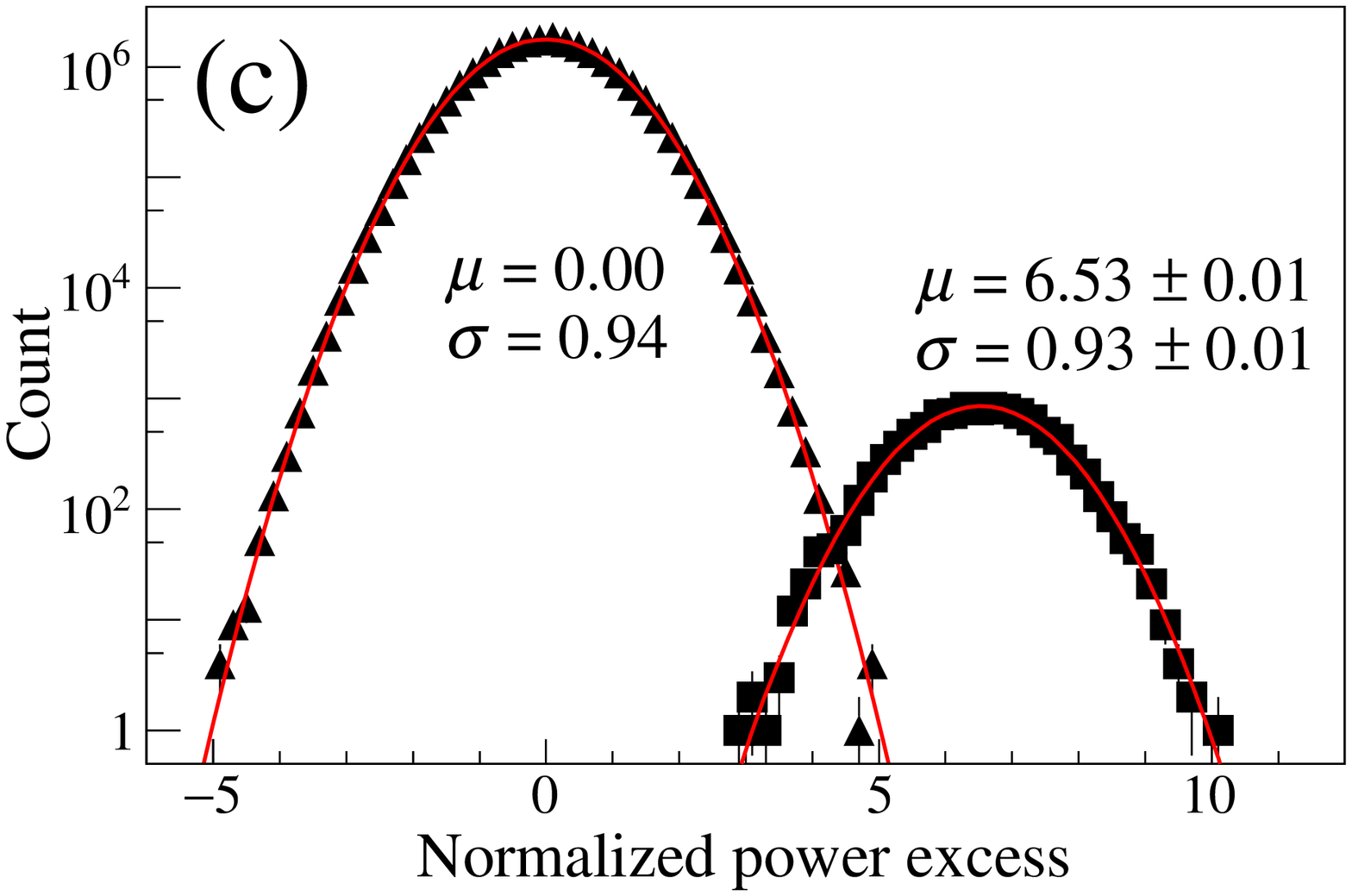}}
  \caption{Circles in (a) show the distribution of the normalized
    power excess from all the frequencies in the normalized grand
    power spectrum from the CAPP-12TB experiment, after applying a
    frequency-independent scale factor of 0.94 which is the original
    width of the distribution.
    (b) shows the distributions of the normalized power excess from
    10000 simulated CAPP-12TB experiments by co-adding 81 adjacent
    50~Hz power spectral lines after the background subtraction using
    the simulation input, and (c) shows those by 9 adjacent 450~Hz
    power spectral lines and the $\chi^2$ fit.    
    Rectangles and triangles in (b) and (c) are from the frequency 
    with the simulated axion signals and frequencies with background
    only, respectively, where both are from the same simulated signals
    with an initial signal-to-noise ratio of 7.89.    
    Lines are a Gaussian fit resulting in $\mu$ (mean) and $\sigma$
    (width).}  
  \label{FIG:CAPP-12TB-PULLS}
\end{figure*}

Our receiver chain consists of a single data acquisition (DAQ)
channel. As shown in Fig.~\ref{FIG:CAPP-12TB-chain}, power from the cavity
goes through a directional coupler, two circulators, and a JPA, which
were located in the magnetic-field cancellation region realized by a
magnet system~\cite{OI}.

The JPA is described and characterized in detail
elsewhere~\cite{CAPP-JPA1,CAPP-JPA2}, and here we describe our JPA
operation scheme for this work.
The JPA profiles were probed with a $+$1 kHz offset from the JPA
resonant frequencies which were set accordingly to the target
frequencies with a $-$100~kHz offset, where the target frequencies are
the central frequencies of each individual power spectra.
The JPA gains $G_{\rm{JPA}}$, detailed in Appendix~\ref{JPA_TUNE},
were measured through the ``Bypass'' line including a 50-$\Omega$
termination (the ``Noise source'' in Fig.~\ref{FIG:CAPP-12TB-chain})
by the microwave switch. They were about 17 dB at the target
frequencies over the frequency range. The noise temperatures of the
JPA $T_{\rm{JPA}}$ were measured to be about 60~mK at the target
frequencies from the power ratio with and without JPA
amplification~\cite{ADMX_DFSZ2} (see also Appendix~\ref{JPA_TUNE}).
The {\it in-situ} $G_{\rm{JPA}}$ through the line including the cavity
instead of the 50-$\Omega$ termination were measured every time the
cavity is tuned during data collection. The power was further
amplified inside the fridge using two serial LNF-LNC0.6\_2A~\cite{LNF}
High-Electron-Mobility Transistor (HEMT) amplifiers anchored at the
4-K stage.

After further processing outside the fridge (see also
Fig.~\ref{FIG:CAPP-12TB}), which comprises of downconversion to the
intermediate frequency (IF) of 10.7~MHz with an image rejection
mixer~\cite{POLYPHASE} and additional amplification, the power was
then digitized and converted into a frequency spectrum over a span
from $-$500 to $+$500~kHz with respect to the IF with a resolution
bandwidth (RBW) of 10~Hz utilizing the fast DAQ system~\cite{FDAQ},
accessed via CULDAQ~\cite{CULDAQ}.

The gain and noise temperature from the receiver chain other than the
JPA, $G_{\rm{JPA_{off}}}$ and $T_{\rm{JPA_{off}}}$, were measured
using the Y-factor method~\cite{YFACTOR} by varying the physical
temperature of the ``Noise source'' up to 400~mK without JPA
amplification.
Accordingly they include all the attenuation and noise from the rest
of the chain and are about 104~dB and 1.2~K, respectively, where the
latter contributes about 25~mK to the noise temperature measurements
with the 50-$\Omega$ termination according to Ref.~\cite{FRIIS} at the
target frequencies. The total gain of the receiver chain including the
{\it in-situ} $G_{\rm{JPA}}$ of about 17 dB, $G_{\rm{total}}$, is
about 121~dB at the target frequencies.

The total system noise temperature $T_n$, detailed in
Appendix~\ref{TOTAL_NOISE}, was obtained from every individual power
spectra by eliminating the $G_{\rm{total}}$ in power and then
parameterized with a Savitzky-Golay (SG) filter~\cite{SGF}.
The parameters of the SG filter were a polynomial of degree 4 and a
2001-point window at an RBW of 10~Hz.
The extracted $T_n$ at the cavity mode $\nu$ is about 215~mK as an
approximately Lorentzian peak (see Fig.~\ref{FIG:CAPP-12TB-TN} in
Appendix~\ref{TOTAL_NOISE}), which is attributed to the tuning rod
being hotter than the cavity walls due to the poor thermal link
between them and the piezoelectric motor operation for the frequency
tuning.
\begin{figure*}
  \centering
  \includegraphics[width=0.85\textwidth]{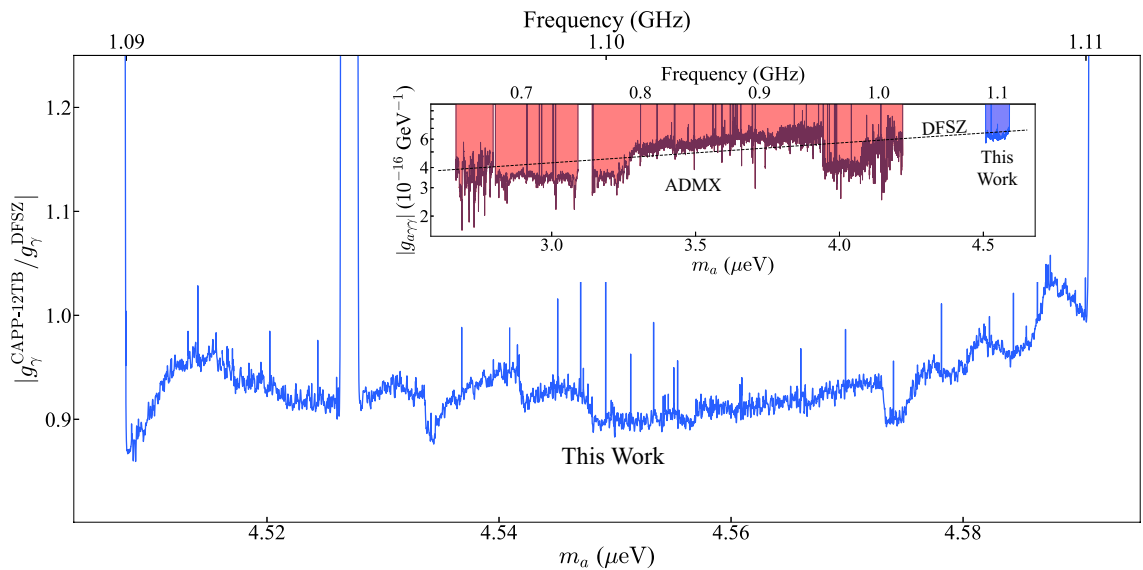}  
  \caption{Blue solid line is the excluded parameter space at a 90\%
    confidence level by this work. Note the mode crossing around
    an axion mass of 4.527 $\mu$eV, which is predicted by a finite
    element method calculation~\cite{CSTCOMSOL}. The inset shows
    exclusion limits from other axion haloscope searches sensitive to
    the DFSZ axion dark matter~\cite{ADMX_DFSZ1, ADMX_DFSZ2} as well
    as that from this work. The intermittent spikes are less sensitive
    due to the filtering procedure (see the text) reducing statistics
    at the given frequency points.}
  \label{FIG:CAPP-12TB-LIMIT}
\end{figure*}

The signal-to-noise ratio (SNR) in this work is defined by the signal
power for the DFSZ axion dark matter coupling $g^{\rm DFSZ}_{a\gamma\gamma}$ 
\begin{equation}
  {\rm SNR}=\frac{P^{g^{\rm DFSZ}_{a\gamma\gamma}}_a}{P_n}\sqrt{b_a\Delta t}=\frac{P^{g^{\rm DFSZ}_{a\gamma\gamma}}_a}{k_B b_a T_n}\sqrt{N}=\frac{P^{g^{\rm DFSZ}_{a\gamma\gamma}}_a}{\sigma_{P_n}}
  \label{EQ:SNR}
\end{equation}
according to the radiometer equation~\cite{DICKE}, where $P_n$ and
$\sigma_{P_n}$ are the total noise power and its fluctuation,
respectively. $b_a$ is the axion signal window, $\Delta t$ is the
integration time at each step, $N$ is the number of
power spectra, and $k_B$ is the Boltzmann constant. We acquired data
from March 1st to 18th in 2022 including system maintenance.
A total of 1996 resonant frequencies were scanned over a search range
of 20.06~MHz with frequency steps of 10~kHz, except for 380~kHz due to
the mode crossings around an axion mass of 4.527~$\mu$eV.
We made sure that the cavity simulation~\cite{CSTCOMSOL} also
observed the mode crossings.
Power spectra were taken with $\Delta t\sim500$ s at each step, which
resulted in $N\sim5000$ with our RBW choice of 10~Hz~\cite{FDAQ}. The
power spectra were averaged and then processed through the analysis
procedure addressed below. The expected SNR values over the search
range were generally higher than 5.

The data analysis process basically follows the axion haloscope
analysis procedures developed
to date~\cite{ADMX_ANAL, HAYSTAC_ANAL, CAPP_ANAL}.
From each power spectrum, only the axion signal sensitive region
around the cavity mode was used. Furthermore, a span of 150~kHz
centered at the target frequency was selected to avoid the presence of
the pump tone and higher noise contribution with lower JPA
gain~\cite{FRIIS}. The frequency span allows 15 power spectra to
overlap in most of the frequency range with our frequency steps of
10~kHz.
First, narrow spikes in each power spectrum were removed with a
filtering procedure similar to the one by the Haloscope At Yale
Sensitive To Axion CDM~\cite{HAYSTAC_ANAL}.
Each power spectrum was parameterized by a $\chi^2$ fit and then went
through the filtering at an RBW of~10 Hz.
The fit function is the product of a five-parameter
function~\cite{ADMX_ANAL} for the overall noise profile of the cavity
and the receiver chain, a Lorentzian for the $G_{\rm{JPA}}$ profile,
and a quadratic or, equivalently, inverse Lorentzian for the noise
profile depending on the $G_{\rm{JPA}}$ profile
(Fig.~\ref{FIG:CAPP-12TB-TN} in Appendix~\ref{TOTAL_NOISE} shows a
hint of this dependence).
In the fit, the minimum of the quadratic function was constrained to
be located at the JPA resonant frequency according to the
$G_{\rm{JPA}}$ profile and the JPA gain at the target frequency was
required to be consistent with the aforementioned measurement.
This $\chi^2$ parametrization improved the SNR efficiency
$\epsilon_{\rm SNR}$ by about 10\% in the end albeit having almost the
same rescan candidates compared to that by an SG filter using the same
parameters mentioned above.
Five nonoverlapping frequency points in each power spectrum were
merged for further filtering at RBW of 50~Hz and the background
subtraction thereafter.
After the background subtraction, each power spectral line and its
fluctuation in each spectrum were scaled by the expected total axion
signal power, where the scaling across the spectrum follows the
Lorentzian line shape characterized by the cavity
$Q_L$~\cite{HAYSTAC_ANAL}.
Allowing the overlaps among the power spectra, all the power spectra
were then combined to produce a single power spectrum and the
associated power fluctuations were also propagated accordingly. The
RBW of the combined spectrum was further reduced to 450~Hz by merging
nine nonoverlapping frequency points, which reduced the SNR as
discussed below. This was, however, a tradeoff to reduce the number of
rescan candidates.
From the combined spectrum, our ``grand power spectrum'' was
constructed by co-adding~\cite{ADMX_ANAL} nine adjacent 450~Hz power
spectral lines. This corresponds to a $b_a$ of 4050 Hz retaining more
than 99.9\% of the putative signal power of axion dark matter
considered in this Letter. Each power spectral line in the grand power
spectrum was weighted by the axion signal shape, a boosted
Maxwellian~\cite{HAYSTAC_ANAL}.
The grand power spectrum was normalized by $\sigma_{P_n}$ which was
also weighted according to the signal shape~\cite{HAYSTAC_ANAL} and
co-added~\cite{ADMX_ANAL}. A frequency-independent scale factor of
0.94 was applied to remedy the bias in power excess induced from the
background subtraction~\cite{CAPP_ANAL}, resulting in a normalized
grand power spectrum following the standard Gaussian as shown in
Fig.~\ref{FIG:CAPP-12TB-PULLS}(a).
With such standard Gaussian statistics, we applied a threshold of
3.718$\sigma_{P_n}$ which is not only to get a one-sided 90\% upper
limit corresponding to the expected axion signal power of
5$\sigma_{P_n}$ from the CAPP-12TB experiment, but also manageable
rescan candidates.
We found 33 power spectral lines located in 14 individual power
spectra exceeded the cut. After rescanning the 14 individual power
spectra with sufficiently high statistics, no power spectral lines
were found to exceed the cut.
\begin{figure}
  \centering
  \includegraphics[width=0.42\textwidth]{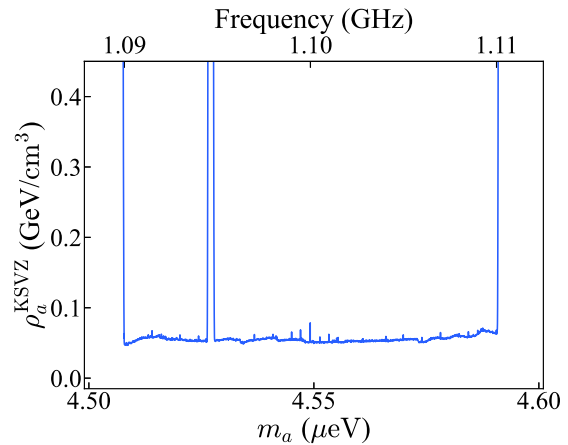}  
  \caption{Exclusion limits for the KSVZ axion dark matter density
    $\rho^{\rm KSVZ}_a$ at a 90\% confidence level.}  
  \label{FIG:CAPP-12TB-LIMIT-RHO}
\end{figure}

The $\epsilon_{\rm SNR}$ was estimated from 10000 simulated CAPP-12TB
experiments with axion signals at a particular frequency on top of
the CAPP-12TB background~\cite{CAPP_ANAL}.
Figures~\ref{FIG:CAPP-12TB-PULLS}(b) and (c) show the normalized power
excess distributions from the simulation, where
Fig.~\ref{FIG:CAPP-12TB-PULLS}(b) was obtained by co-adding 81
adjacent 50~Hz power spectral lines after the background subtraction
using the simulation input, and Fig.~\ref{FIG:CAPP-12TB-PULLS}(c) by 9
adjacent 450~Hz power spectral lines and our $\chi^2$ fit.
The narrow width of the normalized power excess distribution 0.94
induced from the background subtraction shown in the CAPP-12TB data
was also demonstrated by the large statistics simulation data, as
shown in the solid triangles in Fig.~\ref{FIG:CAPP-12TB-PULLS}(c).
The $\epsilon_{\rm SNR}$ at an RBW of 450~Hz with respect to the SNR
at an RBW of 50~Hz is estimated to be 88\% from the means of the
rectangle distributions in Figs.~\ref{FIG:CAPP-12TB-PULLS}(b) and (c),
and also with the frequency-independent scale factor of 0.94.
This 88\% is attributed to not only the signal weighting in the
co-adding procedure with the reduced RBW from 50 to 450~Hz
(95\%)~\cite{RBW_REDUCTION}, but also the background subtraction
(93\%).
The former implies that there is SNR degradation depending on the RBW
choice and the latter is our $\epsilon_{\rm SNR}$ with respect to the
SNR at an RBW of 450~Hz.
Other SNR inefficiencies, from the filtering and the axion signal
misalignment with respect to the RBW of 450~Hz, were negligible
compared to that from the background subtraction~\cite{8TB_PRL}. The
total $\epsilon_{\rm SNR}$ taking into account for the background
subtraction, the additional line attenuation (see
Appendix~\ref{ADD_ATTEN}), and the systematic uncertainty (see
Appendix~\ref{TOTAL_NOISE}) were reflected in our exclusion limits.

We set the 90\% upper limits of $g_{a\gamma\gamma}$ for
$4.51<m_a<4.59$ $\mu$eV. Figure~\ref{FIG:CAPP-12TB-LIMIT} shows the
excluded parameter space at a 90\% confidence level (CL) from the
CAPP-12TB experiment. Using the achieved experimental sensitivity, we
also excluded KSVZ axion dark matter, which makes up 13\% of the
$\rho^{\rm{KSVZ}}_a$, as shown in
Fig.~\ref{FIG:CAPP-12TB-LIMIT-RHO}. Our results are the most sensitive
in the relevant axion mass range to date.

In summary, we report a search for DFSZ axion dark matter using the
CAPP-12TB haloscope. The CAPP-12TB experiment has been pushing all the
experimental parameters to reach state-of-the-art performance. With
such performance, the CAPP-12TB experiment has achieved
$g_{a\gamma\gamma}$ of about $6.2\times10^{-16}$ GeV$^{-1}$ which is
beyond the DFSZ axion dark matter coupling, over the axion mass range
between 4.51 and 4.59~$\mu$eV at a 90\% CL. This is an unprecedented
sensitivity tier in the mass range to date. We expect that CAPP-12TB
as a state-of-the-art axion haloscope will continue sensitive searches
over a wide range of axion masses with high-frequency~\cite{HIGHF} and
high-quality cavity designs~\cite{HIGHQ}.

%\acknowledgments
This work was supported by IBS-R017-D1-2022-a00 and JST ERATO (Grant
No. JPMJER {\bf 1601}). Arjan Ferdinand van Loo was supported by a
JSPS postdoctoral fellowship.
\appendix
\section{JPA Gain and Noise}\label{JPA_TUNE}
The JPA gains were measured with a vector network analyzer~(VNA) as
the power ratio with and without the pump power through the ``Pump''
line in Fig.~\ref{FIG:CAPP-12TB-chain} for the parametric
amplification. Equation~(\ref{EQ:RPOWER}) shows the relation between
the power ratio and the JPA characteristics, $T_{\rm{JPA}}$ and
$G_{\rm{JPA}}$.
\begin{equation}
  \frac{P^{\rm{NS}}_{\rm{JPA_{on}}}}{P^{\rm{NS}}_{\rm{JPA_{off}}}}=\frac{\left(T_{\rm{NS}}+T_{\rm{JPA}}+\frac{T_{\rm{JPA_{off}}}}{G_{\rm{JPA}}}\right)G_{\rm{JPA}}}{T_{\rm{NS}}+T_{\rm{JPA_{off}}}},
  \label{EQ:RPOWER}  
\end{equation}
where $P^{\rm{NS}}_{\rm{JPA_{on}}}$ and $P^{\rm{NS}}_{\rm{JPA_{off}}}$
are the powers transferred from a noise source (NS) with and without
JPA amplification, respectively, $T_{\rm{NS}}$ and
$T_{\rm{JPA_{off}}}$ are the noise temperatures from an NS and the
receiver chain without JPA amplification, respectively. For the JPA
gain measurement, the noise source was the VNA and the power from the
VNA was set much higher than other two power sources, satisfying the
relation
$\frac{P^{\rm{VNA}}_{\rm{JPA_{on}}}}{P^{\rm{VNA}}_{\rm{JPA_{off}}}}\simeq\frac{T_{\rm{VNA}}G_{\rm{JPA}}}{T_{\rm{VNA}}}=G_{\rm{JPA}}$.
We required the JPA gains at the probe tones to be $20\pm0.4$~dB and
thus always obtained gains of around 20~dB at the JPA resonant
frequencies. Our target frequencies had a $+$100~kHz offset from the JPA
resonant frequencies, so the {\it in-situ} $G_{\rm{JPA}}$ for this
axion dark matter search was about 17~dB at the cavity modes over the
frequency range, while those at $-$100 and $+$100~kHz detuned from the
cavity modes were about 20 and 14~dB, respectively, with JPA
bandwidths of about 190~kHz.

The JPA noise temperatures were also measured using the power ratio
with and without the JPA amplification, which also follows
Eq.~(\ref{EQ:RPOWER}).
For the JPA noise temperature measurement, the noise source was the
50-$\Omega$ termination (denoted as ``Noise source'' in
Fig.~\ref{FIG:CAPP-12TB-chain}). The physical temperature of the
50-$\Omega$ termination was maintained at about 25~mK, resulting in a
$T_{\rm{50\mbox{-}\Omega}}$ of about 34~mK for the frequency range
considered in this Letter according to the standard quantum limit.
Hence, one can extract the $T_{\rm{JPA}}$ from Eq.~(\ref{EQ:RPOWER})
using the two independent measurements, the aforementioned
$G_{\rm{JPA}}$ and the $T_{\rm{JPA_{off}}}$ using the Y-factor method
in the text, resulting in $\frac{T_{\rm{JPA_{off}}}}{G_{\rm{JPA}}}$ is
about 25~mK at the target frequencies.

One can also extract the $T_{\rm{JPA}}$ by varying the physical
temperature of the ``Noise source'', i.e., using the Y-factor
method. We found that varying the physical temperature of the ``Noise
source'' also varied $G_{\rm{JPA}}$. This was why we were not able to
apply the Y-factor method to get the reliable $T_{\rm{JPA}}$. However,
varying the physical temperature of the 50-$\Omega$ termination from
100 to 400~mK did not affect the HEMT physical temperature. It enabled
us to apply the Y-factor method for the $T_{\rm{JPA_{off}}}$
measurements.

\section{Total Noise}\label{TOTAL_NOISE}
\begin{figure}[h]
  \centering
  \includegraphics[width=0.42\textwidth]{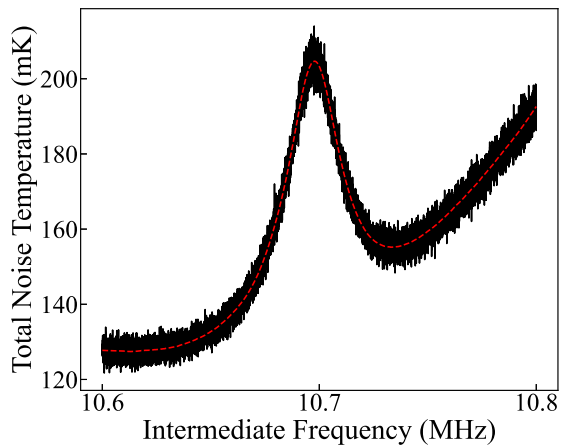}
  \caption{$T_n=\frac{P^{\rm{cavity}}_{\rm{JPA_{on}}}}{k_B{\Delta f}G_{\rm{total}}}$
    as a function of frequency from a particular frequency step. Red
    dashed lines represents the SG filter parametrization of $T_n$.}  
  \label{FIG:CAPP-12TB-TN}
\end{figure}

Equation~(\ref{EQ:TNOISE}) shows the relation between the measured
power transferred from the cavity, the total system noise temperature
$T_n$, and the total system gain $G_{\rm{total}}$.
\begin{equation}
  P^{\rm{cavity}}_{\rm{JPA_{on}}}=k_B{\Delta f} T_n G_{\rm{total}},
  \label{EQ:TNOISE}
\end{equation}
where $\Delta f$ is the RBW. Figure~\ref{FIG:CAPP-12TB-TN} shows a
typical $T_n$ as a function of frequency downconverted to the IF at a
particular frequency step, after eliminating the $G_{\rm{total}}$ from
the measured power $P^{\rm{cavity}}_{\rm{JPA_{on}}}$.
As shown in Fig.~\ref{FIG:CAPP-12TB-TN}, the total noise temperatures
at $-$100 and $+$100~kHz detuned from the target frequency were
typically 130 and 190~mK, which mainly comes from the Lorentzian JPA
gain profile that determines the noise temperature profiles of the
chain after the JPA~\cite{FRIIS} and of the JPA itself
(see Eq.~(\ref{EQ:RPOWER})) as well. Considering the noise profile
comes from the Lorentzian JPA gain profile, the $T_n$ around the
target frequency is about 140~mK. A na\"ive estimate of the $T_n$ from
the aforementioned measurements with the ``Noise source'' is about
120~mK which is a sum of the noise contributions shown in the
parenthesis in Eq.~(\ref{EQ:RPOWER}), where they are 34, 60, and
25~mK, respectively. In this case, the noise source is the cavity.
We admit that there is a 20~mK difference from the physical
temperature difference between the cavity top and bottom due to the
incomplete thermalization between the cavity walls and the hot tuning
rod, where the two temperatures are 25 and 45~mK, respectively.
Provided the cavity noise temperature on average is about 10~mK higher
than that from the cavity top, the 20~mK difference can roughly be
understood, since this additional noise temperature contributes also
to the JPA input referred noise at the idler frequency~\cite{JPAIEEE}.
Nevertheless, our total noise temperature measurement is inclusive and
does not distinguish the three contributions shown in the
parenthesis in Eq.~(\ref{EQ:RPOWER}). Therefore, our resulting
sensitivity is reliable, in spite of the lack of the exact cavity
temperature information, with an accurate $G_{\rm{total}}$ corrected
in Appendix~\ref{ADD_ATTEN}.
In our inclusive $T_n$ measurement, the uncertainty of our temperature
sensor RX-C102B~\cite{LAKESHORE} for the 50-$\Omega$ termination as
the ``Noise source'' in Fig.~\ref{FIG:CAPP-12TB-chain} can contribute
to the ``Noise source'' temperatures in
$G_{\rm{JPA_{off}}}=\frac{P_h-P_c}{k_B\Delta f(T_h-T_c)}$, where $P_h$
and $P_c$ are the powers transferred from the ``Noise source'' whose
temperatures are $T_h$ (400~mK) and $T_c$ (100~mK), respectively,
without the JPA amplification, thus affecting $G_{\rm{total}}$.
The magnetic field dependent temperature sensor error is expected to
be negligible~\cite{LAKESHORE}, because it was located in the
magnetic-field cancellation region realized by the magnet
system~\cite{OI}. The relevant data are limited down to a temperature
of $\mathcal{O}$(1~K), but show errors less than 1\% at 1~T below
independent of the physical temperature of the
sensor~\cite{LAKESHORE}. Nonetheless, we conservatively assigned the
sensor error of 2\% which propagated to about 3\% to
$G_{\rm{JPA_{off}}}$, subsequently to $T_n$.

With the cavity as a noise source, the $T_n$ was also measured using
Eq.~(\ref{EQ:RPOWER}),
$\frac{P^{\rm{cavity}}_{\rm{JPA_{on}}}}{P^{\rm{cavity}}_{\rm{JPA_{off}}}}=\frac{T_n G_{\rm{JPA}}}{T_{\rm{cavity}}+T_{\rm{JPA_{off}}}}$.
Here, the $T_n$ was extracted inclusively with two other measurements,
the aforementioned $G_{\rm{JPA}}$ and the $T_{\rm{JPA_{off}}}$ from
the Y-factor method in the text. We found the difference between the
two noise measurements was at most 5\% resulted from our imperfect
measurements. Otherwise, they are likely consistent with each other
because the two measurements are almost fully dependent.

The total systematic uncertainty of our $T_n$ measurement was
estimated to be 6\% by combining the two sources above.

\section{Additional Line Attenuation}\label{ADD_ATTEN}
The line attenuation from the cavity to the microwave switch was not
considered in our total gain $G_{\rm{total}}$ mentioned in the
text. Using a VNA and ``Strong'' line in
Fig.~\ref{FIG:CAPP-12TB-chain}, we measured two reflected powers, one
from the cavity and the other from the switch. In order to obtain the
fully reflected power, the switch was configured to float between the
two switch ports and the cavity off-resonant frequency range was
inspected. The line attenuation was measured by taking half of the
difference between the two reflection measurements and was 0.3--0.5~dB
depending on the frequency.


\begin{thebibliography}{99}
\bibitem{RBW_REDUCTION}  
  Note also that the RBW reduction from 10 to 50~Hz decreases the SNR less than 0.1\%.
\end{thebibliography}
\end{document}